\documentclass[fleqn,usenatbib]{mnras}

\usepackage{mathptmx}
\usepackage[T1]{fontenc}
\usepackage{ae,aecompl}

\usepackage{graphicx}	% Including figure files
\usepackage{amsmath}	% Advanced maths commands
\usepackage{amssymb}	% Extra maths symbols
\usepackage{commath}

\usepackage[detect-all,alsoload=astro,separate-uncertainty=true]{siunitx}
\usepackage[all]{hypcap}
\usepackage{xspace}
\usepackage{newtxtext,newtxmath}
\hypersetup{breaklinks=true}
\newcommand{\hmpc}{\ensuremath{h^{-1}\mathrm{Mpc}}\xspace}

\newcommand{\abacus}{A\textsc{bacus}\xspace}
\newcommand{\abacuscosmos}{A\textsc{bacus} C\textsc{osmos}\xspace}
\newcommand{\rockstar}{R\textsc{ockstar}\xspace}
\newcommand{\astropy}{Astropy\xspace}
\newcommand{\halotools}{\texttt{Halotools}\xspace}

\title[Tests of Acoustic Scale Shifts in Halo-based Mocks]{Tests of Acoustic Scale Shifts in Halo-based Mock Galaxy Catalogues}

\author[Duan Y. T. et al.]{
	Duan Yutong,$^{1}$\thanks{E-mail: dyt@physics.bu.edu}
	Daniel Eisenstein,$^{2}$
	\\
	$^{1}$Physics Department, Boston University, 590 Commonwealth Ave, Boston, MA 02215, USA\\
	$^{2}$Institute for Theory and Computation, Harvard-Smithsonian Center for Astrophysics, 60 Garden St, Cambridge, MA 02138, USA
}

\date{Accepted XXX. Received YYY; in original form ZZZ}

\pubyear{2019}

\begin{document}
	\label{firstpage}
	\pagerange{\pageref{firstpage}--\pageref{lastpage}}
	\maketitle
	
	\begin{abstract}
	
		We utilise mock catalogues from high-accuracy cosmological $N$-body simulations to quantify shifts in the recovery of the acoustic scale that could potentially result from galaxy clustering bias. The relationship between galaxies and dark matter halos presents a complicated source of systematic errors in modern redshift surveys, particularly when aiming to make cosmological measurements to sub-percent precision. Apart from a scalar, linear bias parameter accounting for the density contrast ratio between matter tracers and the true matter distribution, other types of galaxy bias, such as assembly and velocity biases, may also significantly alter clustering signals from small to large scales. We create mocks based on generalised halo occupation populations of 36 periodic boxes from the \abacuscosmos release, and test various biased models along with an unbiased base case in a total volume of $48 \,  h^{-3} \si{\giga\parsec\tothe{3}}$. Two reconstruction methods are applied to galaxy samples and the apparent acoustic scale is derived by fitting the two-point correlation function multipoles. With respect to the baseline, we find a 0.3\% shift in the line-of-sight acoustic scale for one variation in the satellite galaxy population, and we find a 0.7\% shift for an extreme level of velocity bias of the central galaxies. All other bias models are consistent with zero shift at the 0.2\% level after reconstruction. We note that the bias models explored are relatively large variations, producing sizeable and likely distinguishable changes in small-scale clustering, the modelling of which would further calibrate the BAO standard ruler.
		
	\end{abstract}
	
	
	%%%%%%%%%%%%%%%%%%%%%%%%%%%%%%%%%%%%%%%%%%%%%%%%%%
	
	%%%%%%%%%%%%%%%%% BODY OF PAPER %%%%%%%%%%%%%%%%%%

	\section{Introduction}
	
	The standard ruler provided by baryon acoustic oscillations (BAO) has become a powerful probe in the past decade for studying the large-scale structure of the universe and constraining properties of dark energy. The next generation of dark energy experiments, such as the Dark Energy Spectroscopic Instrument (DESI) \citep{desi_exp_pt1}, Euclid \citep{euclid}, and WFIRST \citep{wfirst}, are designed and built with BAO as a primary method to measure the expansion history of the universe to unprecedented precision. The acoustic scale around 100\hmpc is much larger than the scales relevant for nonlinear gravitational evolution, and galaxy formation and remains in the linear regime today. However, nonlinear effects still have small but important consequences, despite the acoustic scale being a robust standard ruler for measuring cosmological distances.  Systematic errors on the distance scales inferred from BAO measurements are indeed dominated by nonlinear structure growth as well as redshift-space distortions. As reconstruction has been shown to substantially reduce these major contributors to systematics if not reverse them entirely \citep{sherwin2012}, the next subleading source of systematics, galaxy bias, is increasingly relevant with the precision of BAO measurements reaching the $0.1\%$ level in future surveys \citep{mehta_galaxy_bias}.
	
	Density field reconstruction uses Lagrangian perturbation theory to reduce anisotropies in the clustering and reverse the large-scale gravitational bulk flows in the observed galaxy sample \citep{recon_std}. This procedure is very effective in undoing the smoothing of the BAO feature and the shift of the BAO scale due to nonlinear structure growth, thereby improving the precision of BAO measurements. The one-step version, known as standard reconstruction, has been field tested extensively with on-sky data \citep{2percent, anderson2012, anderson_dr9, anderson_dr11, wigglez, boss_dr12_bao} and become part of the standard analysis procedure in modern galaxy redshift surveys. There has been much recent development in reconstruction \citep{2017JCAP...12..009S, 2017PhRvD..96b3505S, 2018PhRvD..97b3505S, 2018PhRvD..97d3502Z, recon_ite, 2019ApJ...870..116W}. While standard reconstruction has worked well to restore and enhance the BAO signature, whether newer algorithms can perform better is a question. To find out, we employ one of the new methods, the iterative reconstruction in particular \citep{recon_ite}, which provides better correlation to the matter density field and closer-to-truth statistics than standard reconstruction when applied to galaxy mocks \citep{hada_application}.
	
	Galaxies as tracers of the underlying matter density field are far from perfect and tend to overweigh overdense regions. It is well known that galaxy bias can be scale-dependent and, even with a linear bias correction, shift the acoustic scale measurement \citep{2009PhRvD..80f3508P, mehta_galaxy_bias}, increasing systematic errors in cosmic distances and cosmological parameters. \citet{mehta_galaxy_bias} showed that the one-step reconstruction \citep{recon_std} reduces the BAO shifts due to linear bias $b\leqslant 3.29$ and results in no significant shift of the BAO scalein a volume of $44 \, h^{-3}\si{\giga\parsec\tothe{3}}$. Yet, many mechanisms in the halo model can give rise to galaxy bias that may not be well approximated as a linear parameter, and can induce sub-percent level shifts of the acoustic scale. For example, galaxy or halo assembly bias is a significant source of systematic error in the galaxy-halo relationship at small scales \citep{zentner_assembly_bias}. The CMASS BOSS galaxy sample was found to have significant velocity bias for the central galaxies, and between velocity and spatial distributions of satellites, at least one is biased \citep{2015MNRAS.446..578G}. \citet{wu13} found significant changes of the power spectrum caused by non-Poisson distribution and velocity bias of satellite galaxies.
	
	This study seeks to assess the sensitivity of the acoustic scale in the two-point correlation functions (2PCF) to the effects of galaxy bias. Our analysis packages together a number of improvements over existing literature. The anisotropic acoustic scale is measured in both radial and transverse directions. Several types of galaxy bias are considered, including satellite distribution bias, assembly bias for central and satellite galaxies, velocity bias for centrals and satellites, and more sub-halo-scale effects. The cosmological $N$-body simulations are generated by the high-accuracy \abacus code \citep{abacus_ic, abacus_compare}, and the total simulation volume of $48 \, h^{-3}\si{\giga\parsec\tothe{3}}$ is the largest as of the writing of this paper. A newer iterative reconstruction method is applied in addition to standard reconstruction \citep{recon_ite}. Halo catalogues are populated with mock galaxies in a generalised halo occupation distribution (HOD) model akin to \citet{yuan_hod}, which can be implemented in a deterministic way such that clustering statistics are differentiable with respect to variations of the input HOD model parameters.

	The layout of this paper is as follows. In \S\ref{sec:methods} methods, \S\ref{sec:sim_halocat} describes the $N$-body simulations and halo catalogues, \S\ref{sec:galcat} covers generation of mock galaxy catalogues with biased HOD models and reconstruction parameters, \S\ref{sec:jk_2pcf}\textendash \ref{sec:cov} goes through the correlation and covariance calculation, and \S\ref{sec:bao_fitting} motivates the BAO fitting test for iterative reconstruction with alternative models. In \S\ref{sec:results}, we examine the differential change in 2PCF resulting from all bias models in \S\ref{sec:2pcf_biases}, and then presents the shift of the BAO scale measurement for of each type of bias in \S\ref{sec:alpha_biases}. Finally, the conclusions section \S\ref{sec:conclusions} highlights the findings.
	
	\section{Simulations, Catalogues, and Methods}
	\label{sec:methods}
	
	\subsection{Simulations and Halo Catalogues}
	\label{sec:sim_halocat}
	
	Our study uses cosmological N-body simulations produced by the \abacus code \citep{abacus_compare}. Specifically, 36 simulation boxes assuming the Planck cosmology \citep{planck15} are used. They all have particle mass about $4\times10^{10} \, h^{-1}M_\odot$, particle count $1440^3$, box size 1100\hmpc, periodic boundary conditions, and initial conditions created with the same input linear power spectrum and independent initial phases \citep{abacus_ic}. The total volume adds up to $48 \, h^{-3}\si{\giga\parsec\tothe{3}}$. The only difference among the 36 boxes is that 16 were run with Plummer force softening and the other 20 with spline softening. Detailed descriptions of the force softening can be found in the \abacus Cosmos public data release paper \citep{abacus_cosmos}. These two sets are first analysed separately from applying halo mass cuts all the way through to the BAO scale measurements, where we find all clustering statistics within $1\sigma$ range of each other when analysing matter density field and galaxy samples created by any given HOD model. There is no difference between two simulations at any confidence level, despite their slightly different halo mass functions. With fixed halo finder and HOD parameters, spline softening typically results in more massive and large halos, meaning more halos pass the mass cut and more galaxies are generated. But as we have found, the BAO scale evolves slowly with respect to the halo mass function, and it is safe to combined two simulations for a larger total volume. All results hereafter have two sets combined and treated as one simulation of 36 boxes.
	
	Several time slices of the simulations after $z=1$ are saved on disk with data products available. We take the $z=0.5$ snapshot to mimic luminous red galaxy (LRG) samples and to stay close to the galaxy redshifts in the BOSS DR12 dataset \citep{2015ApJS..219...12A} as well as the mocks used to analyse it \citep{2017MNRAS.470.2617A}. The halo catalogues are created by the \rockstar halo finder \citep{rockstar}. To incorporate galaxy assembly bias later in the HOD, the halo NFW concentration, defined as $c_\text{NFW} \equiv R_\text{virial}/R_\text{s, Klypin}$, is added to the halo properties, where the virial radius of the halo is chosen as $r_\text{virial} = r_\text{200}$ and the scale radius is given by \citet{klypin_scale}, which is more stable than the traditional scale radius for small halos \citep{rockstar}.
	
	A mass cut at 70 dark matter (DM) particles, or approximately $M_\text{halo} = 4\times 10^{12} M_\odot$, was applied, as small halos have essentially zero chance of hosting galaxies and slow down the computation. Subhalos are not reliable indicators of of in-halo galaxy distribution and are removed as well. Instead, we use the position and velocity of dark matter particles to generate satellite galaxies. Each halo catalogue is accompanied by a DM particle catalogue, which is a 10\% subsample of the particles enclosed by halo boundaries. Particles within subhalos are associated only to their host subhalos in \rockstar catalogues, not the higher level host halos. This becomes a problem after subhalos are removed, and the particle associations have to be rebuilt before satellite generation to ensure that all particles are associated to their highest-level host halos. In addition, a $10\%$ uniform subsample of all $1440^3$ particles in any given simulation box was also used to validate our BAO fitter, as well as to quantify how much cosmic variance there exists in the BAO measurements and are cancelled out when the difference is taken.
	
	\subsection{Galaxy and Random Catalogues}
	\label{sec:galcat}
	
	\subsubsection{Halo Occupation Distribution}
	\label{sec:hod}
	
	The HOD model used for populating halos with galaxies is based on the classic 5-parameter model by \citet{zheng07} with decorations accounting for assembly bias, velocity bias, satellite distribution bias, and perihelion distance bias \citep{yuan_hod}. \abacus provides direct access to DM particles from which halos have been found in the simulation. Although halo finders only produce spherical halo boundaries and therefore spherical DM particle distributions, the matter distribution is still a much better representation of the actual density profile within a halo than NFW profiles. Satellite galaxies generated with DM particles more realistically trace the matter distribution of the halo.
	
	A key feature of this HOD model is the deterministically seeded random numbers used for populating each simulation box. The seed is chosen such that all halos and DM particles always receive the same random number assignment completely irrespective of the HOD parameters specified. This means any infinitesimal change in the input HOD parameters would correspond to an infinitesimal change in the clustering results.
	
	Our revamped HOD implementation\footnote{Publicly available at  \url{https://github.com/duanyutong/abacus_baofit}. This repository also includes our new BAO fitter described in \S\ref{sec:bao_fitting}} largely shares the same formalism and equations as GRAND-HOD \citep{yuan_hod}, so we focus on only the differences or advantages it offers in terms of science and software implementation. The mean halo occupations for central and satellite galaxies take the following forms,
	\begin{align}
	\left< N_\text{cen} (M) \right> &= \frac{1}{2} \left[ 1 + \mathrm{erf} \left( \frac{\ln M - \ln M_\text{cut}}{\sqrt{2}\sigma^\prime} \right) \right] \nonumber \\
	&= \frac{1}{2} \mathrm{erfc} \left( \frac{\ln(M_\text{cut} / M)}{\sqrt{2}\sigma^\prime} \right) \label{eqn:N_cen}\\
	\left< N_\text{sat} (M) \right> &= \left< N_\text{cen} (M) \right> \left( \frac{M - \kappa M _\text{cut} }{M_1} \right) ^\alpha \label{eqn:N_sat}\, .
	\end{align}
	This parametrisation is consistent with the original HOD prescription by \citet{zheng07}, up to a rescaling of the dispersion $\sigma^\prime \equiv (\ln10\sqrt{2}) \sigma$ for a more natural interpretation \citep{white11} and $\kappa \equiv M_0/M_\text{cut}$, and is a popular choice in recent literature \citep{2016JCAP...05..051S, 2016MNRAS.458.4015Z, 2016MNRAS.463.1929M, 2017ApJ...846...61G, 2017ApJ...848...76Z, 2018MNRAS.478.1042S, 2018MNRAS.480.3177B}. The explicit coupling between $\left< N_\text{sat} \right>$ to $\left< N_\text{cen} \right>$ by multiplication maintains the reasonably physical assumption that the central and satellite galaxies in the same halo are correlated to some extent, and that a high-mass halo hosting already hosting a central galaxy has a higher chance of also hosting one or more satellites. Accordingly, we make satellite occupation terminate at a higher halo mass scale than the central occupation cutoff. The dependence of $\left< N_\text{sat} (M) \right>$ on $\left< N_\text{cen} (M) \right>$ introduces complications in the fitting procedures when one tries to determine the coupled central and satellite parameters simultaneously in a given HOD model. If the goal is not to constrain HOD parameters and this fitting difficulty is not of concern, then there is no advantage or motivation for dropping this assumption and insisting on no correlation between centrals and satellites \citep{2017MNRAS.465.2833C}.
	
	Assembly bias is implemented as comparing the halo concentration to the median concentration of all halos of that mass. The more concentrated halo may have a more favourable assembly history and a higher probability of hosting central or satellite galaxies, or vice versa. The median halo concentration as a function of halo mass $c_\text{med}(M)$ is obtained by putting all halos into mass bins and fitting a polynomial to $c_\text{med}(M)$. Instead of taking the halos from a single simulation box as the sample and repeating the fit for all boxes, we take the entire halo population from all boxes in a given simulation and perform the fitting once for all. This has a better theoretical motivation because the function $c_\text{med}(M)$, in principle, is independent from the phase of initial conditions and does not vary across boxes. Even though the entire halo population is now an order of magnitude larger than that of a single box, if a smooth fit is desired and the bin specification is fine, there are still some mass bins with few halos and thereby, small variances in halo concentration. The usual weight definition for all mass bins is $w=1/\sigma_c = \sqrt{(N-1)/\Sigma_i (c_i-\mu_c)}$ up to a normalisation constant, where $N$ is the number of halos in the mass bin, $c_i$ is the concentration of each halo in the bin, and $\mu_c$ is the mean concentration of the bin. With $\sqrt{N}$ in the numerator, this gives disproportionally large weights for those least populated bins, resulting in poor fits. We corrected for this pathology by adjusting the bin weight definition, multiplying the canonical weight by $(N-1)$ in powers of $1/2$ in an attempt to increase the weight for the more populated bins. Both $w=\sqrt{N-1}/\sigma_c$ and $w=(N-1)/\sigma_c$ produced quality fits which were almost identical. We chose $w \equiv \sqrt{N-1}/\sigma_c = (N-1)/\sqrt{\Sigma_i (c_i-\mu_c)}$ as the weight definition.
	
	To optimise I/O performance and the size of data products, \abacus saves each halo catalogue for a single box into many HDF5 files, each being a subset of the halo catalogue. When populating halos, GRAND-HOD proceeds on a subset-by-subset basis as it goes through the HDF5 files sequentially. As a result, all halo ranking operations involved in the decorations, e.g. pseudomass calculations, are limited to the current HDF5 subset of halos only. Our code utilises the halo reader built-in to \abacus and loads the complete halo catalogue at once in the standard \halotools \citep{halotools} format together with the DM particle subsample with corrected host IDs as part of the mock. The new code fully conforms to the \astropy \citep{astropy} and \halotools standards, supports the \halotools prebuilt HOD models, and is compatible with both the latest Python 3 and 2 builds. It provides flexibility in customising HOD parameters and preset models, and boosts performance with parallelism.
	
	\subsubsection{Biased HOD Models}
	\label{sec:biased hod}
	
	\begin{table*}
		\caption{Parameters of the biased HOD models that are tested. The first five columns are input for the classic 5-parameter vanilla HOD model as defined in Eqn. \ref{eqn:N_cen} and \ref{eqn:N_sat}. The last six decorations parameters are briefly reviewed in \textsection\ref{sec:biased hod}. All model parameters are given relatively extreme values intentionally. Dash line means no change from the baseline model. The satellite parameters in Base 2 and Base 3 models are tuned to maintain a constant galaxy number density in the simulation box with respect to the baseline.}
		\label{tab:hod_param}
		\begin{tabular}{lrrrrrrrrrrr}
			\hline
			\hline
			Biased HOD Models & $\log_{10} M_\text{cut}$ & $\sigma^\prime$ & $\kappa$ & $\log_{10} M_1$ & $\alpha$ & $A_\text{cen}$ & $A_\text{sat}$ & $ \alpha_\text{cen}$ & $s$ & $s_v$ & $s_p$ \\
			\hline
			Base 1 (Baseline)
			& 13.35 & 0.85 & 1.0 & 13.800 & 1.00 & 0 & 0 & 0 & 0 & 0 & 0 \\
			Base 2
			& - & - & - & 13.770 & 0.75 & - & - & - & - & - & - \\
			Base 3
			& - & - & - & 13.848 & 1.25 & - & - & - & - & - & - \\
			Assembly Bias (Centrals $+$)
			& - & - & - & - & - & $1.0$ & - & - & - & - & - \\
			Assembly Bias (Centrals $-$)
			& - & - & - & - & - & $-1.0$ & - & - & - & - & - \\
			Assembly Bias (Satellites $+$)
			& - & - & - & - & - & - & $1.0$  & - & - & - & - \\
			Assembly Bias (Satellites $-$)
			& - & - & - & - & - & - & $-1.0$ & - & - & - & - \\
			Velocity Bias (Centrals 20\%)
			& - & - & - & - & - & - & - & 0.2 & - & - & - \\
			Velocity Bias (Centrals 100\%)
			& - & - & - & - & - & - & - & 1.0 & - & - & - \\
			Halo Centric Distance Bias (Satellites $+$)
			& - & - & - & - & - & - & - & - & $0.9$ & - & - \\
			Halo Centric Distance Bias (Satellites $-$)
			& - & - & - & - & - & - & - & - & $-0.9$ & - & - \\
			Velocity Bias (Satellites $+$)
			& - & - & - & - & - & - & - & - & - & $0.9$ & - \\
			Velocity Bias (Satellites $-$)
			& - & - & - & - & - & - & - & - & - & $-0.9$ & - \\
			Perihelion Distance Bias (Satellites $+$)
			& - & - & - & - & - & - & - & - & - & - & $0.9$ \\
			Perihelion Distance Bias (Satellites $-$)
			& - & - & - & - & - & - & - & - & - & - & $-0.9$ \\
			\hline
		\end{tabular}
	\end{table*}
	The generalised HOD framework easily allows any aforementioned source of bias, or combination of sources, to be introduced into the model with flexibility. The biased HOD models tested in this paper are summarised in Table \ref{tab:hod_param}. All models, except the baseline (denoted Base 1), have only \textit{a single} mechanism of galaxy bias applied in order to explore the ``unit vector'' directions in the space of variations. Although this HOD parametrisation does not \textit{exactly} preserve the number density of galaxies, the resulting number density of all biased models only differ from the baseline by about $0.1\%$, which is negligible. By default, central galaxies inherit the velocity of the host halos, and satellite galaxies generated with DM particles assume the DM particle velocities, unless velocity bias is applied. To understand how each bias mechanism changes clustering statistics and the acoustic scale, we first establish a base case, free of any decoration for reference. The baseline is a simple 5-parameter model; its parameters are unimportant as we are only interested in the differential change with respect to the baseline. We reiterate that our HOD implementation is \textit{differentiable}, which is precisely what enables us to set up a baseline and subtract it from the bias model results. For each simulation box, the random numbers are strictly model-independent with both the order and quantity being fixed.
	
	Besides the baseline two more undecorated models are defined, which
	vary only the vanilla HOD parameters. The models named Base 2 and 3 differ from the baseline only in the satellite parameters in Eqn. \ref{eqn:N_sat}: the exponent $\alpha$ is changed by $\pm 25\%$, and the denominator in the power law term, $M_1$, is tuned accordingly. The specific combinations of $M_1$ and $\alpha$ values are chosen to match the number density in the baseline model, $4\times 10^{-4} h^{3}\si{\mega\parsec\tothe{-3}}$, which agrees with realistic LRG sample densities.
	
	Next, there are a number of single-decoration models that come in pairs. Single-decoration means that each model has only one origin of bias present with respect to the baseline model. And each pair of models have opposite changes in the parameter(s) of interest. Two models have assembly bias only for central galaxies with opposite assembly bias parameter $A_\text{cen}$, and similarly two assembly bias models for satellites only. Then there are two central velocity bias models, one assumes a more realistic, $20\percent \cdot v_\text{rms}$ velocity dispersion for the central galaxy relative to the DM halo, and the other assuming a more extreme, $100\percent \cdot  v_\text{rms}$ dispersion. The last six models investigate three bias effects arising from sub-halo-scale astrophysics, by giving preferential treatment to DM particles based on the particle's speed, halo centric distance, or total mechanical energy (quantified by perihelion distance) when assigning satellite galaxies. Below is a brief review of the definitions of the parameters, and readers are referred to \citet{yuan_hod} for a more detailed discussion.
	
	In Table \ref{tab:hod_param}, the first five columns are the standard 5 parameters as defined in Eqn. \ref{eqn:N_cen} and \ref{eqn:N_sat} which govern the mean halo occupation for central and satellite galaxies. $A_\text{cen}$ and $A_\text{sat}$ are assembly bias parameters for centrals and satellites, implemented as comparing the halo concentration to its peers of similar masses and re-assigning
	\begin{equation}
	\log M_\text{pseudo} = \log M + A_\text{cen/sat} \left[ 2 \Theta(c-c_\text{med}) - 1 \right],
	\end{equation}
	a pseudomass to the halo as input for $\langle N_\text{cen}(M) \rangle$ or $\langle N_\text{sat}(M) \rangle$. Here $c_\text{med}$ is the median concentration in the halo mass bin to which the halo belongs, and $\Theta(c-c_\text{med})$ is the Heaviside step function. While \citet{croton07} showed that the formation time (redshift) and halo concentration do not capture the assembly history of halos as far as small-scale two-halo terms in the correlation function is concerned, we still use halo concentration as a proxy for assembly bias, as our mock central and satellite galaxies and do not exactly preserve the 1-halo terms and bias effects may manifest at large scales.
	
	The velocity bias for central galaxies draws randomly from a normal distribution whose width is scaled by $\alpha_\text{cen}$, and adds that peculiar velocity relative to the DM halo on top of the line-of-sight component of the halo velocity,
	\begin{align}
	v_{\varparallel \text{pec}} &\sim N(0, \alpha_\text{cen} \frac{v_\text{rms}}{\sqrt{3}}) \\
	v_\varparallel ^\prime &= v_\varparallel + v_{\varparallel \text{pec}}
	\end{align}
	where $v_\text{rms}$ is the RMS velocity dispersion of the DM particles within the halo. It is known that central galaxies are not at rest relative to their host halos, and their velocity dispersions have been estimated to be of order 10\% of that of the halos in observed galaxies (including LRGs) and in simulations: \citet{2015MNRAS.453.4368G, 2015MNRAS.446..578G} found $\alpha_\text{cen} \approx 0.3$ in BOSS DR7 and $\alpha_\text{cen} = 0.22^{+0.03}_{-0.04}$ in BOSS DR11 using HOD models built on $N$-body simulations, and \citet{2017ApJ...841...45Y} had $\alpha_\text{cen} \gtrsim 0.04$ in hydrodynamical simulations from the Illustris suite. The uncertainty is usually a few percent and $\alpha_c$ depends on how the halo reference frame is exactly defined, but the consensus on central velocity dispersion is 10\% to 30\%. In light of these results, we choose $\alpha_\text{cen} = 20\%$ as a realistic case and also test an extreme level of central velocity bias in another model where $\alpha_\text{cen} = 100\%$, which is implausible but could be more revealing of this particular bias effect on the clustering statistics.
	
	The last three $s$ parameters control in-halo satellite generation by modifying the probability of each particle hosting a satellite as
	\begin{equation}
	p_i = \overline{p} \left[ 1 + s_{\_,v,p} (1 - \frac{2 r_i}{N_\text{part} - 1}) \right],
	\end{equation}
	where $\overline{p} \equiv \langle N_\text{sat} (M) \rangle / N_\text{part}$ is the uniform probability for each particle to begin with, the three ranking parameters need to satisfy $s_{\_,v,p} \in (-1, 1)$ to conserve the total probability, $N_\text{part}$ is the total number of particles within the halo, and $r_i=0, 1, 2, \ldots, N_\text{part}-1$ is $i$th particle's ranking by halo centric distance, speed, or perihelion (total mechanical energy), all at the snapshot taken at $z=0.5$.
	
	We stress that the parameters for the bias models listed in Table \ref{tab:hod_param} are intentionally chosen to be quite extreme. For a common 20-particle halo and $s=0.9$, for example, the innermost particle (rank $r=0$)  has 19 times the probability of the outermost particle ($r=19$) to match to a satellite galaxy. The purpose is to increase the chance of detecting a shift in the acoustic peak location and to explore the worse-case scenarios.
	
	\subsubsection{Redshift-Space Distortion and Reconstruction}
	\label{sec:rsd_recon}
	
	By adding decorations to the base HOD class, we are modifying the line-of-sight velocity of galaxies and assuming an alternative truth velocity. As apparent RSD depends on the true peculiar velocity in addition to the Hubble flow, we artificially apply RSD to the line-of-sight coordinate as the last step of mock galaxy generation by modifying $x^\prime_\varparallel = x_\varparallel + v_\varparallel / \left[a H(a) \right]$ in the line-of-sight direction, after all decorations are completed. In practice, this is implemented as $x^\prime_\varparallel = x_\varparallel + v_\varparallel / \left[a H_0 E(z) \right]$ where $E(z)$ is the \astropy \texttt{efunc} defined as $H(z) \equiv H_0 E(z)$.
	
	Now the complete mock galaxy catalogue is ready to be treated as observed data. Two reconstruction methods are applied and compared side by side: the standard reconstruction \citep{recon_std} and a recent iterative reconstruction method \citep{recon_ite}. For both reconstruction methods, the optimal smoothing scale $\Sigma = 15\hmpc$ is used. Additional parameters for iterative reconstruction used are grid size $N_\text{grid} = 480^3$, galaxy bias $b = 2.23$, initial smoothing scale $\Sigma_\text{ini} = 15\hmpc$, annealing parameter $\mathcal{D} = 1.2$, weight $w = 0.7$, and number of iterations $n_\text{iter} = 6$. These parameters are chosen based on \citet{recon_ite, hada_application} where variations in the input galaxy bias $b$ up to $20\percent$ was found to hardly impact the iterative reconstruction result.
	
	The rest of the procedures including 2PCF, covariance, and fitting is performed on all three types of galaxy catalogues: pre-reconstruction, post-reconstruction (standard), and post-reconstruction (iterative).
	
	\subsection{Noise Suppression in Two-Point Correlation Functions}
	\label{sec:jk_2pcf}
	
	To suppress the shot noise in galaxy generation, for the same simulation box and same biased HOD model, 12 realisations of the galaxy catalogue are generated repeatedly with varied initial seed. The initial condition phase for a box $p$ is an integer index labelling the simulation box. The realisation index $r$ is an integer from 0 to 11. The random number generator seed is reset as $s = 100p + r$ before every galaxy generation, which guarantees that the random numbers are always deterministic and model-independent, as long as we never exceed 100 realisations. Tn the beginning of each galaxy catalogue generation, a fixed quantity of random numbers ($N_\text{halos} + N_\text{particles}$) are thrown for centrals and satellites before any other operation takes place which may involve throwing more random numbers (e.g. central velocity bias which draws randomly from a Gaussian distribution). As $N_\text{halos}$ and $N_\text{particles}$ are both constant for a given simulation box, this ensures that the same random numbers are always generated, regardless of the HOD model imposed, and assigned to every halo or DM particle in a fixed order.
	
	For each realisation of the galaxy sample, the 2PCF and their Legendre multipole decompositions are calculated using a Fast Fourier Transform algorithm on the $k$-grid \citep{fftcorr}, as well as using the pair-counting method for cross-checking when applicable. All pair-counting is done in fine $(s, \mu)$ and $(r_p, \pi)$ bins using a highly efficient pair-counting code Corrfunc \citep{2017ascl.soft03003S, 10.1007/978-981-13-7729-7_1}: $s$ bin edges from 0 to 150\hmpc at 1\hmpc steps and $\mu$ bin edges from 0 to 1 at 0.01 steps. The pair-counts are re-binned with optimal bin size $\Delta s = 5\hmpc$ found in \citet{boss_dr12_bao}.
	
	2PCF functions are calculated from raw pair-counts using a generalised form of the \citet{ls_estimator} estimator, which works for both auto- and cross-correlations. Given data samples $D_1, D_2$, random samples $R_1, R_2$ in the same respective volumes, and sample sizes $N_{D1}, N_{D2}, N_{R1}, N_{R2}$ (number of data or random galaxies in the sample), the correlation as a function of pair-counts is
	\begin{align}
	\label{eqn:ls_estimator}
	\xi_\text{LS}
	&= \frac{\frac{D_1D_2}{N_{D1}N_{D2}} - \frac{D_1R_2}{N_{D1}N_{R2}} - \frac{R_1D_2}{N_{R1}N_{D2}} + \frac{R_1R_2}{N_{R1}N_{R2}}}{\frac{R_1R_2}{N_{R1}N_{R2}}} \nonumber\\
	&= \frac{\overline{D_1D_2} - \overline{D_1R_2} - \overline{R_1D_2} + \overline{R_1R_2}}{\overline{R_1R_2}}
	\end{align}
	where pair-counts with bars denotes normalised pair-counts, i.e. raw pair-counts weighted by sample population sizes $1/(N_1N_2)$. For the auto-correlation of galaxy samples, we may simply set $D_1 = D_2$ and $R_1 = R_2$. No FKP weighting is included as the galaxy distribution is homogeneous in one redshift bin in our simulations.
	
	\begin{figure*}
		\includegraphics[width=\linewidth]{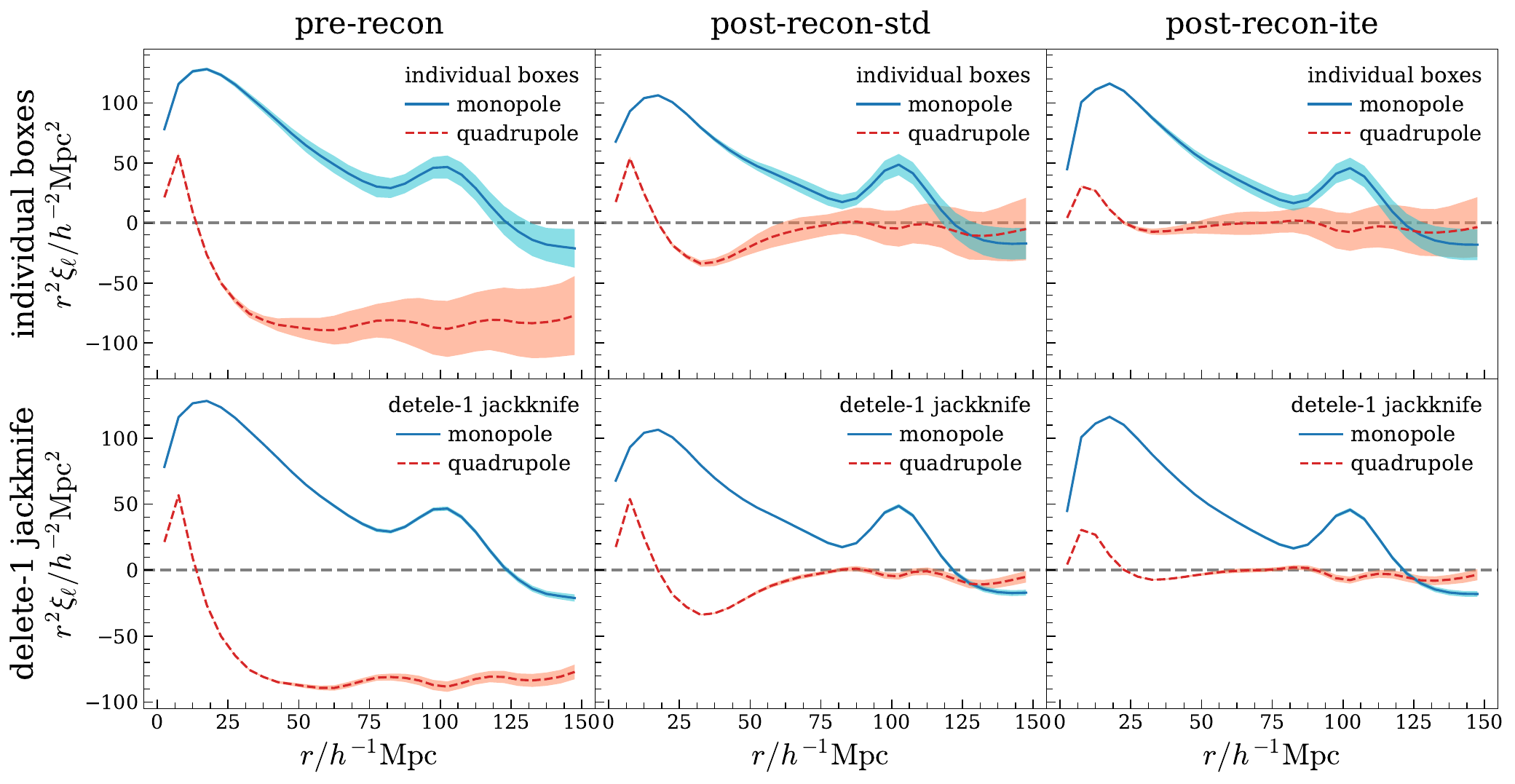}
		\caption{
			(Colour online) The first two multipoles of the galaxy 2PCF for the baseline HOD model showing fluctuations among simulation boxes. The first row shows $N_\text{box}=36$ multipole samples derived from co-adding 12 realisations of each box; the curves are the mean monopole or quadrupole, and the shaded regions are $\pm1\sigma$ intervals around the mean. The second row shows the re-sampled, delete-1 jackknife multipoles derived from co-adding $N_\text{box}-1$ boxes at a time; the shaded $\pm1\sigma$ regions are in fact the standard deviation rescaled by $\sqrt{N_\text{box}-1}=\sqrt{35}$ to account for the jackknife re-sampling, though difficult to see. Three columns are pre-reconstruction, post-reconstruction (standard), and post-reconstruction (iterative). From left to right, the monopole BAO peak is sharpened and the quadrupole getting closer to zero, indicating less anisotropy and better restored spherical BAO shell. Although different boxes in a given simulation only differ by the initial condition phase and over ten realisations are generated and co-added, there are still considerable fluctuations in the 2PCF among boxes. Jackknife re-sampling yields much smoother and stabler samples, and greatly reduces the uncertainty in the mean correlations.}
		\label{fig:phase_jackknife}
	\end{figure*}
	The multipoles from all realisations are co-added into one correlation sample for a given simulation box. To further increase signal-to-noise ratio in the clustering statistics, we take the delete-1 jackknife samples of the multipoles by ignoring one box and co-adding all the other boxes at a time. A comparison between individual box samples and jackknife samples is shown in Fig. \ref{fig:phase_jackknife}. There are large enough fluctuations across different boxes that jackknife re-sampling is a necessary step before BAO fitting and significantly reduces sample variance. The $N_\text{box}=36$ jackknife samples are then passed on to the fitter.
	
	\subsection{Covariance Estimation}
	\label{sec:cov}
	
	The covariance between all multipoles and $(s, \mu)$ bins in the 2PCF must be estimated before fitting for any physical parameter. Since we are interested in the acoustic scale around 100\hmpc, an order of magnitude smaller than the simulation box size 1100\hmpc, we opt to divide the box into subvolumes, increasing the number of correlation samples while still retaining the BAO signal, and bootstrap the covariance. We choose $N_\text{sub} = 3$ along each dimension, so that each subvolume has a side length of over 3 times the BAO scale of interest. For each realisation, the full box galaxy sample is divided into $N_\text{sub}^3 = 27$ subvolumes. The galaxies in each subvolume are cross-correlated with the full box volume using Eqn. \ref{eqn:ls_estimator}, with index 1 being the full box and index 2 being the subvolume. Every subvolume is treated independently, and all its realisations are co-added to prevent shot noise from entering the covariance matrix.
	
	By taking cross-correlations between the full box and the subvolume, we obtain $N_\text{sub}^3=27$ times the number of auto-correlation samples in $1/N_\text{sub}^3=1/27$ of the box volume. In the end, the joint monopole-quadrupole covariance matrix is derived from $N_\text{box}N_\text{sub}^3=36\times 3^3=972$ correlation samples. As covariance scales inversely with the spatial volume in which it is calculated, and the auto-correlations used in the fitting are for the full box, this covariance matrix is re-scaled by a factor of $1/N_\text{sub}^3$ to account for the subvolume division. An additional factor of $1/(N_\text{box}-1)$ is needed if fitting to jackknife multipoles averaged over $N_\text{box}-1$ samples.
	
	It is worth noting that for every bias model and every type of correlation there is a different covariance matrix. Pre-reconstruction galaxy samples are given uniform, analytic randoms to calculate the correlations and covariance. Standard reconstruction produces a shifted galaxy catalogue as well as shifted numerical randoms, which can then be both subdivided to estimate the covariance. Our standard reconstruction implementation produces a shifted random set 200 times the size of the data set. For the purpose of estimating covariance, it is computationally expensive and unnecessary to use all of it. We opt to speed up the pair-counting by randomly downsampling the random set to a level of 10 times the data. To further balance the pair-counting workload between the $DR$ and $RR$ terms in the correlation estimator for covariance bootstrap, while $D_1R_2$ and $R_1D_2$ are counted using the $10\times$ subsample of shifted randoms, for $R_1R_2$ the $10\times$ subsample is split into 10 copies, each of size $1\times$ the galaxy sample, and $R_1R_2$ is counted 10 times using the $1\times$ split samples and then the pair-counts averaged. Iterative reconstruction does not provide any shifted sample after it completes, only the auto-correlations. We assume it shares the same covariance matrix as standard reconstruction, given how similar their post-reconstruction correlations are.
	
	For BAO fitting, which involves about 10 degree of freedoms, this estimate of the covariance matrix is acceptable but of course not perfect \citep{2014MNRAS.439.2531P}. We emphasise that the purpose is to determine the shifts in the acoustic scale, not the confidence level of the chi-square fit in the $(\alpha_\perp, \alpha_\varparallel)$ space. The covariance matrices only weight the fit overall and still give the correct acoustic scale.
	
	\subsection{Fitting 2PCF for the BAO Scale}
	\label{sec:bao_fitting}
	
	\begin{figure*}
		\includegraphics[width=\linewidth]{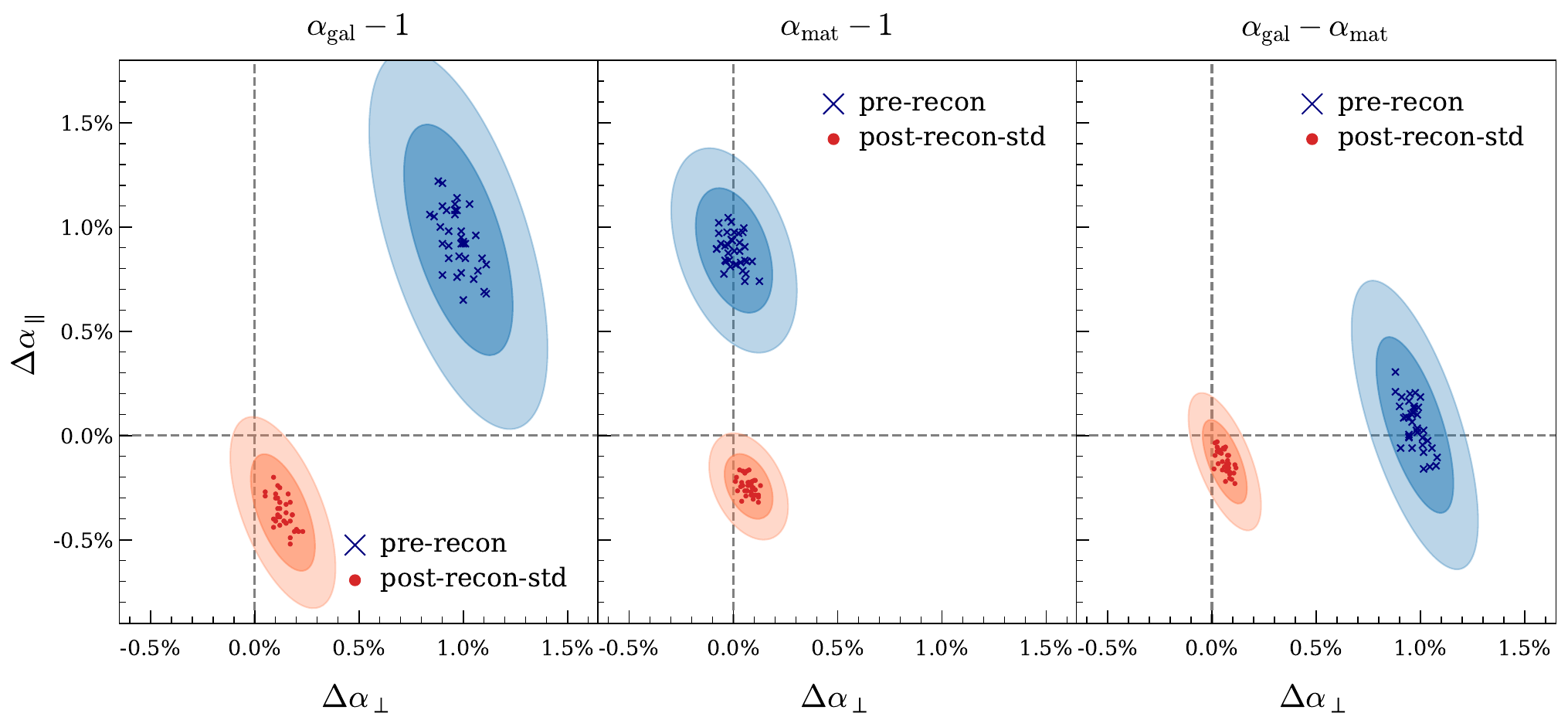}
		\caption{
			(Colour online) BAO fitting results for uniform matter density filed and the mock galaxy sample in the baseline HOD model. From left to right, the first panel shows baseline mock galaxy BAO scale deviation from 1 in the transverse and radial directions, the second panel shows the same for matter density field, and the third panel shows their difference $\alpha_\text{gal} - \alpha_\text{mat}$. In each panel, the data points are derived from fitting $N_\text{box} = 36$ jackknife samples, and the shaded $1\sigma, 2\sigma$ confidence regions are scaled accordingly to reflect the true uncertainty. Comparing panels 1 and 3, the variance visibly decreases when subtracting out the matter field result from the galaxy result, and the mean residue is close to the origin post-reconstruction, indicating galaxy bias in the base model has introduced no statistically significant shift relative to the matter.
			\label{fig:sample_variance_cancel}}
	\end{figure*}
	Following the tried-and-true fitting methods described in previous BAO analyses of galaxy redshift surveys \citep{anderson_dr9, anderson_dr11, boss_dr12_bao}, we fit to the monopole and quadrupole jackknife samples and determine the anisotropic acoustic scale. This is done by performing a $\chi^2$ grid scan in the radial and transverse acoustic scale plane $(\alpha_\varparallel, \alpha_\perp)$, marginalising over all polynomial nuisance parameters in the monopole and quadrupole templates. The fiducial fitting model assumes fitting range $r \in (50\hmpc, 150\hmpc)$, bin size 5\hmpc, and the \textit{poly}3 nuisance form (third-order inverse polynomial in power $k$), $A_\ell(r) = A_{2}/r^2 + A_{1}/r + A_{0}$. 
	
	% new fitter, P to xi
	Our new BAO fitter supports any arbitrary input linear power spectrum and transforms it with flexible choices of parameters into correlation multipole templates. Starting with the input linear power spectrum of the simulation (e.g. from C\textsc{amb}) and the no-wiggle power spectrum \citep{EH98},
	\begin{equation}
	P(k, \mu) = C^2(k, \mu, \Sigma_s)\left[ (P_\text{lin} - P_\text{nw}) e^{-k^2\sigma_v^2} + P_\text{nw} \right] \label{eqn:power_template}
	\end{equation}
	where
	\begin{align}
	\sigma_v^2 &=\frac{(1-\mu^2)\Sigma_\perp^2}{2} + \frac{\mu^2 \Sigma_\varparallel^2}{2}\\
	C(k, \mu, \Sigma_s) &= \frac{1+\mu^2\beta\left[ 1-S(k)\right]}{1+\frac{k^2\mu^2\Sigma_s^2}{2}}\\
	S(k) &= e^{-\frac{k^2\Sigma_r^2}{2}}\\
	\Sigma_\varparallel &= \frac{\Sigma_\perp}{1-\beta}.
	\end{align}
	Here the reconstruction smoothing scale $\Sigma_r = 15\hmpc$ and the streaming scale $\Sigma_s = 4\hmpc$ are fixed. The last equation is a convenient approximation such that the user only needs to specify $\Sigma_\perp$, and in the isotropic case, $\beta=0$ enforces $\Sigma_\perp = \Sigma_\varparallel$. We have experimented with various choices of the other parameters and checked which one(s) best recovered the truth acoustic scale in the input power spectrum. For pre-reconstruction matter density field, $\Sigma_\perp = 1.5\hmpc$, and for pre-reconstruction galaxy catalogue, $\Sigma_\perp = 5\hmpc$ provide the appropriate smoothing of the BAO peak. For all post-reconstruction samples, even with the lowest choice $\Sigma_\perp = 0$ and least smoothing, the monopole peak of the template is still slightly wider than that of the data, so $\Sigma_\perp = 0$ is used. When varying $\beta\in [0, 0.5]$, we find that for pre-reconstruction samples, the resulting $\alpha$ scale is relatively stable and insensitive to $\beta$ while $\chi^2$ would increase by up to $50\%$ as $\beta$ increases; for post-reconstruction samples, $\beta=0$ best recovers the truth acoustic scale. Therefore in all cases we set $\beta=0$.
	
	The power spectrum in Eqn. \ref{eqn:power_template} is decomposed into power multipoles and then Fourier transformed to correlation multipoles in the usual manner
	\begin{align}
	P_\ell (k) &= \frac{2\ell + 1}{2} \int_{-1}^{1} P(k, \mu) L_\ell(\mu) \dif \mu \\
	\xi_\ell(r) &= \frac{i^\ell}{2\pi^2} \int_{k_\text{min}}^{k_\text{max}} k^2 P_\ell(k)j_\ell(kr) e^{-(ka)^2 r} \dif k
	\end{align}
	where $L_\ell (\mu)$ is the Legendre polynomial, $j_\ell (kr)$ is the spherical Bessel function of the first kind, $k_\text{min}$ and $k_\text{max}$ are the limits in the input linear power, and $a = 0.35\hmpc$ controls the Gaussian damping term which suppresses high-$k$ oscillations of the Bessel kernel and we found necessary in order to produce the correct shape of correlations multipoles.
	
	% fitter validation, panels 1 and 2
	As a validation of the fitter, we fit to a $0.2\%$ uniform subsample of the matter field ($6\times10^6$ of $1440^3$ DM particles in a box) without RSD applied in addition to mock galaxy samples. We find the fitted matter field acoustic scale in excellent agreement with the theoretical template at the $0.3\%$ level up to nonlinear corrections after standard reconstruction (middle panel of Fig. \ref{fig:sample_variance_cancel}). This level of shift is expected due to nonlinear evolutions when fitting with a linear power spectrum as input \citep{seo08}, and is observed in the first two panels of Fig. \ref{fig:sample_variance_cancel}: the baseline galaxy and uniform matter field fits have mean BAO shifts of $0.3\%$ to $0.4\%$. 

	% cancellation, panel 3
	We also find that $N_\text{box}=36$ jackknife fitting results in the $(\alpha_\varparallel, \alpha_\perp)$ plane are distributed like an ellipse, indicating that the errors are Gaussian and we may estimate the confidence region using the elliptical distribution of points. Ultimately we are concerned with the differential change in the acoustic scale $\alpha$ when a certain source of galaxy bias is introduced, and the sample variance in the fitted $\alpha$ values should largely cancel out. In the last panel of Fig. \ref{fig:sample_variance_cancel} showing the $\alpha_\text{gal} - \alpha_\text{mat}$ subtraction, the uncertainty regions indeed shrink and the data points are much more tightly bound together compared to the first panel without subtraction. For a given simulation box, this subtraction does cancel out a significant amount of sample variance. The post-reconstruction point being at the origin also indicates that galaxy bias in the base model has introduced no statistically significant shift relative to the matter.
	
	\begin{figure}
		\includegraphics[width=\columnwidth]{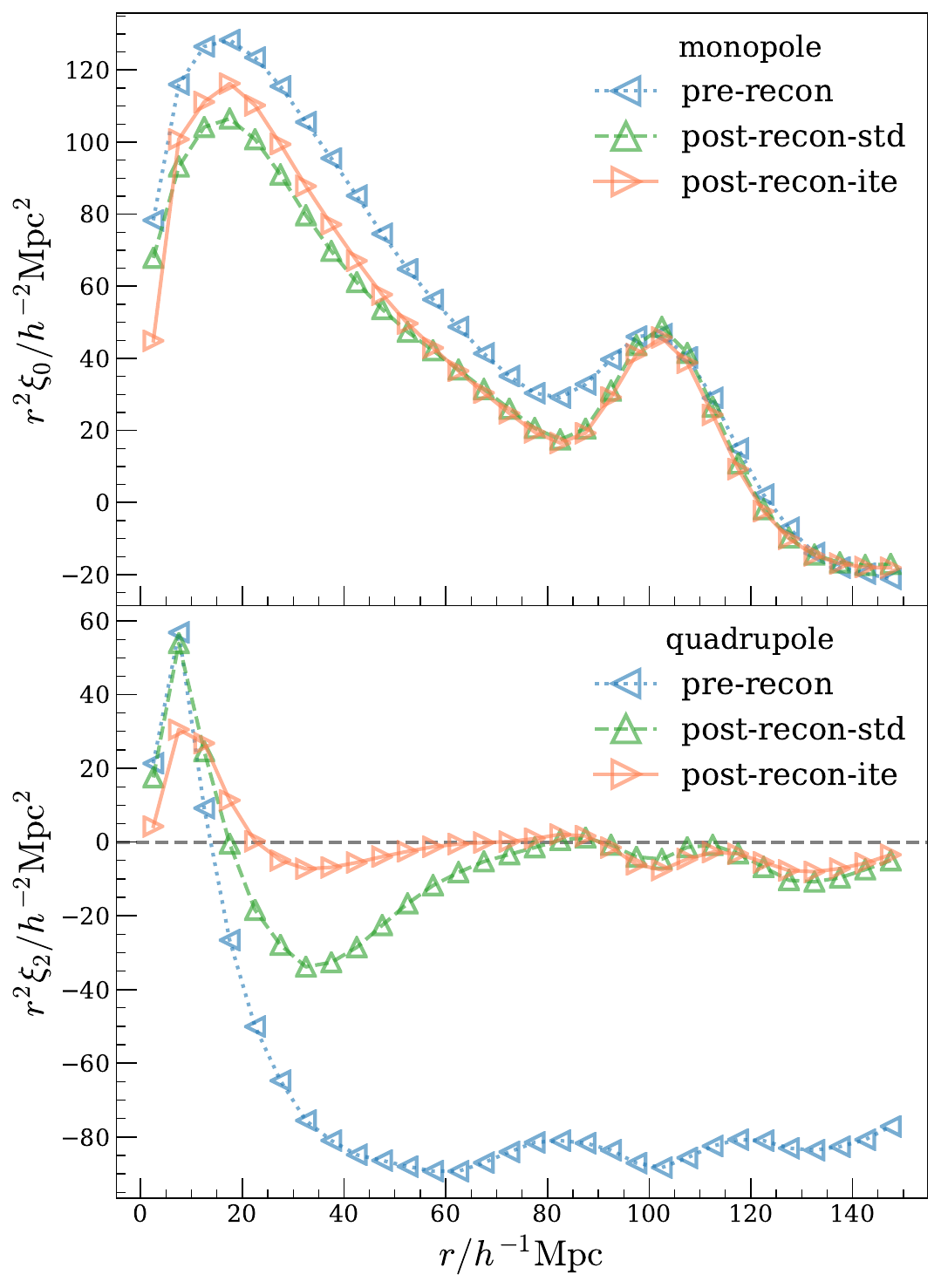}
		\caption{
			(Colour online) An overplotted comparison between pre- and post-reconstruction multipoles for the baseline HOD model, averaged over all simulation boxes. In the top monopole panel, two reconstruction methods both sharpen and narrow the BAO peak, and essentially overlap with each other from 60\hmpc to 150\hmpc, only differing on small-scales. In the bottom quadrupole plot, iterative reconstruction does a significantly better job reducing anisotropy and appears nearly flat down to 20\hmpc, whereas standard reconstruction has large residual anisotropies below 80\hmpc. Two reconstruction methods also produce slightly different monopole on small scales below 50\hmpc, with iterative reconstruction supposedly be more accurate, but this range is usually discarded in BAO fitting to better isolate the BAO signal.}
		\label{fig:tpcf_recon}
	\end{figure}
	
	% tests of fiducial fitting model
	Although the \textit{poly}3 fitting form has been shown to be robust for samples with and without standard reconstruction in previous analyses \citep{2percent, anderson_dr9}, alternative choices are worth considering again for our \abacus mocks with the new iterative reconstruction method applied. Iterative reconstruction does significantly better than standard reconstruction in reducing the anisotropies in the quadrupole, especially on intermediate to small scales, as shown in Fig. \ref{fig:tpcf_recon}. Using a simpler $A(r)$ form or a wider $r$ range in the fitting model might yield better fits. Polynomials with simple Fourier transformation properties are motivated by the need to marginalise over the broadband shape of the galaxy correlation functions and to isolate the BAO feature. Having a nonzero $A(r)$ is important for ameliorating inaccuracies of the assumed fiducial cosmology and keeping the fit robust against variations in the input, and polynomials of degrees as high as 4 risks over-fitting the data with too much freedom. This means that $A_\ell(r) = 0$ (\textit{poly}0) and $A_\ell(r) = A_1/r^2 + A_2/r + A_3 + A_4 r$ (\textit{poly}4) are both disfavoured. The broadband correlation contains unwanted information such as scale-dependent bias, uneven galaxy number densities, and redshift-space distortions among other observational effects, which might not be present in our simulation mocks in the first place. We experimented with several other nuisance forms,
	\begin{align}
	A_\ell(r) &= \frac{A_1}{r^2}						&& (\textit{poly}1)\\
	A_\ell(r) &= \frac{A_1}{r^2} + \frac{A_2}{r}		&& (\textit{poly}2) \\
	A_\ell(r) &= \frac{A_1}{r^2} + \frac{A_2}{r} + A_3	&& (\textit{poly}3) \\
	A_\ell(r) &= \frac{A_2}{r} + A_3 \, .				&& (\textit{poly}3^\prime)
	\end{align}
	along with increased fitting ranges and found no improvement over the fiducial fitting model in any case, in terms of recovering the true BAO $\alpha$ and reducing $\chi^2$ of the fit. The HOD bias results in \textsection\ref{sec:results} are all obtained with the fiducial fitting model.
	
	\section{Results}
	\label{sec:results}
	
	\subsection{Two-Point Correlation Functions of Bias Models}
	\label{sec:2pcf_biases}
	
	% rppi pre/post-recon
	\begin{figure*}
		\includegraphics[width=\linewidth]{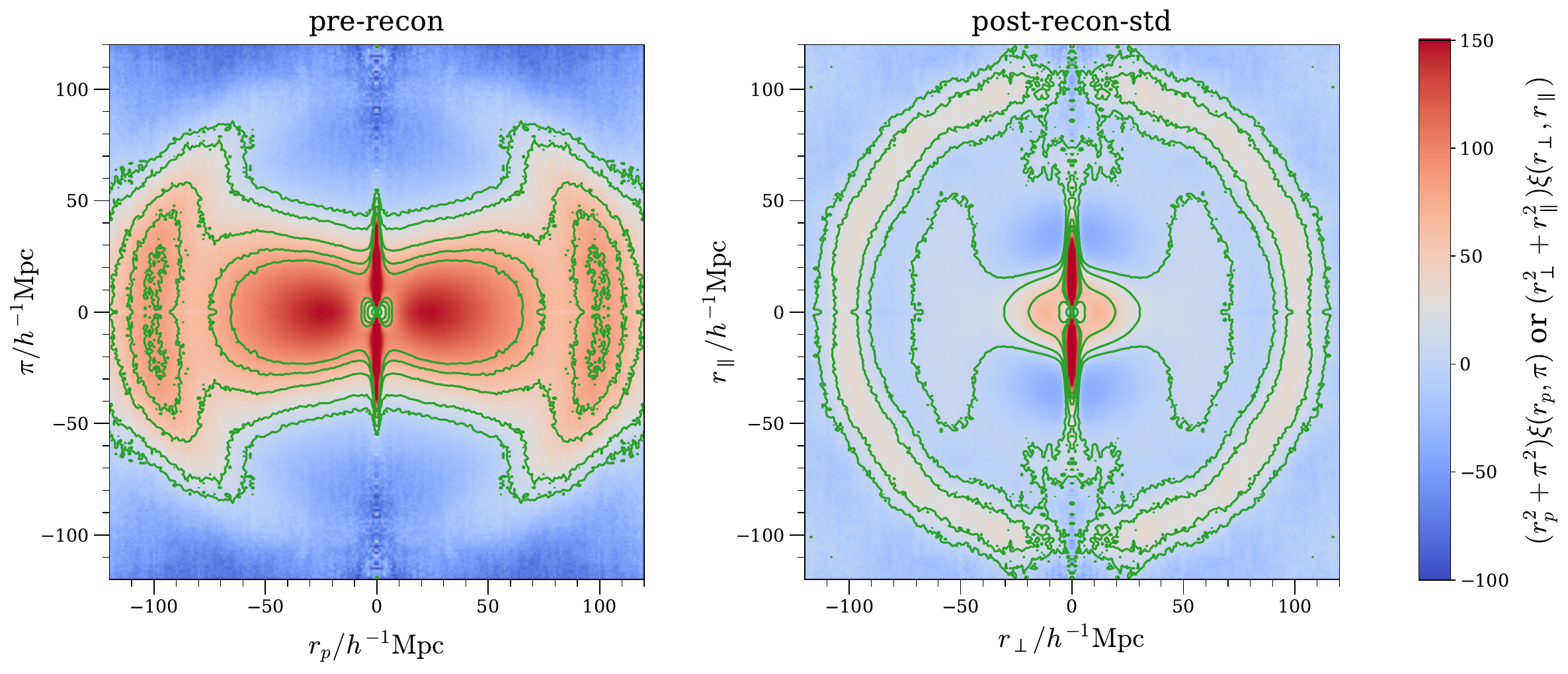}
		\caption{
			(Colour online) Comparison between pre- and post-reconstruction 2D 2PCF for the baseline HOD model, averaged over all boxes, normalised by $r^2$. The left panel is pre-reconstruction, and the right panel is after standard reconstruction is applied. Iso-correlation contours are drawn in green. The 1-step standard reconstruction works well in restoring the spherical BAO shell around 100\hmpc. Other bias models are visually the same in this plot, and there are interesting differences on small scales around the origin which are shown in Fig. \ref{fig:delta_rppi}.}
		\label{fig:rppi_recon}
	\end{figure*}
	Before presenting BAO fitting results, we first examine the 2PCF resulting from different biased HOD models and reconstruction methods. Fig. \ref{fig:rppi_recon} shows the $\xi(r_p, \pi)$ 2PCF for the baseline model before and after standard reconstruction, averaged over all boxes up to 120\hmpc (heatmap colour is rescaled by $r^2 = r_p^2 + \pi^2$ or $r^2 = r_\perp^2 + r_\varparallel^2$). Standard reconstruction does a fine job restoring the isotropy of the spherical BAO shell in the redshift-space 2PCF around $r = 100\hmpc$. On intermediate to large scales, all bias models look very similar in this $\xi(r_p, \pi)$ plot, with differences being obvious only on the small scales.
	
	\begin{figure*}
		\includegraphics[scale=0.43]{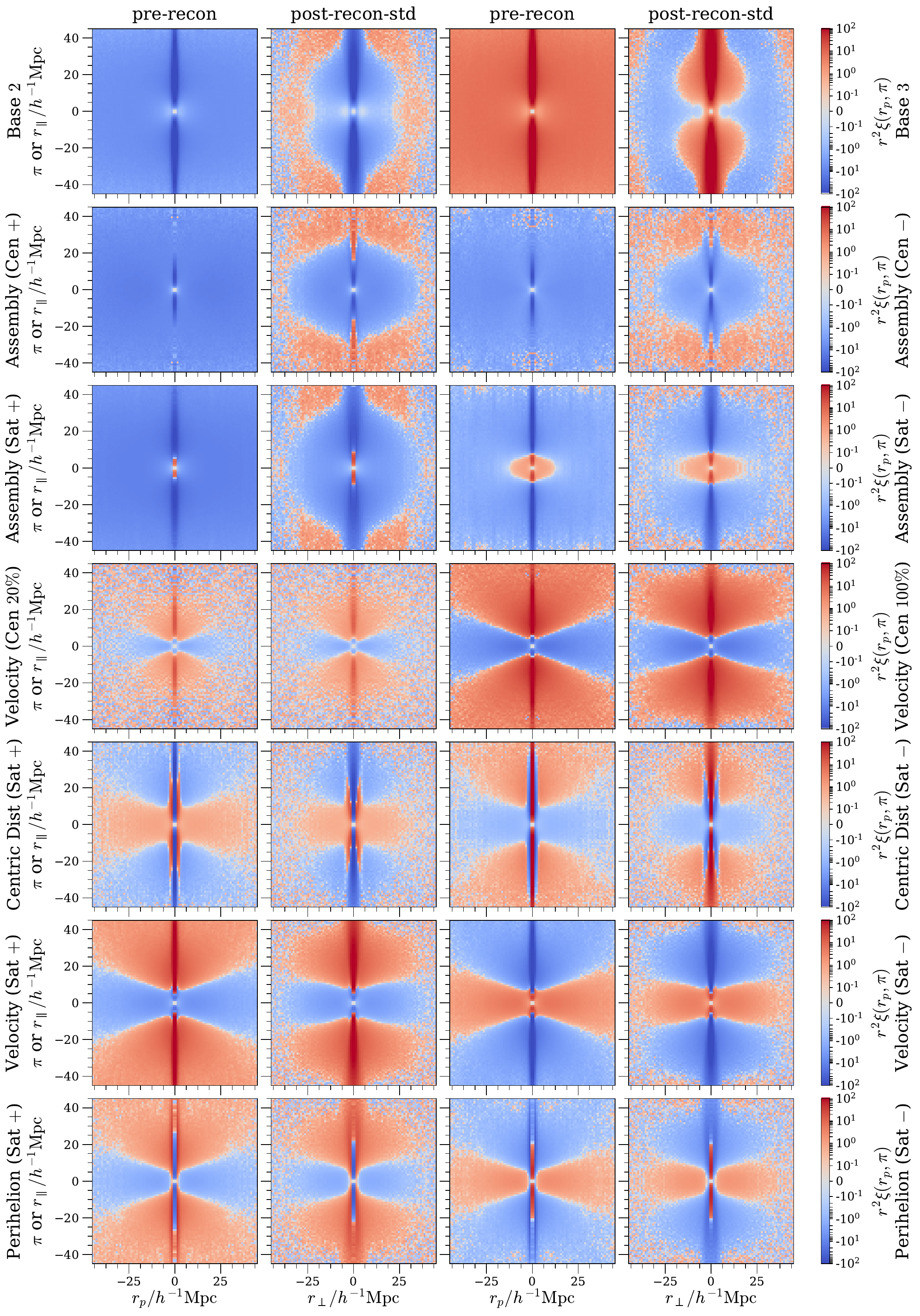}
		\caption{
			(Colour online) Changes in 2D 2PCF of each biased HOD model with respect to the baseline model, averaged over all boxes. Each row contains a pair of two comparable models; each model has a pre-reconstruction panel and a post-reconstruction (standard) one. Symmetrical positive and negative changes in the HOD parameters in all pairs of models but assembly bias ones induce symmetrical change in the correlation $\Delta \xi$ in rows 1, 5, 6 and 7.}
		\label{fig:delta_rppi}
	\end{figure*}
	Zooming in to the small-scale correlations around the origin in Fig. \ref{fig:rppi_recon}, we takes a closer look at the effects of galaxy bias on clustering in Fig. \ref{fig:delta_rppi} by plotting the correlation difference with respect to the baseline model, $\Delta\xi=\xi_\text{model} - \xi_\text{base}$, again rescaled by $r^2 = r_p^2 + \pi^2$ or $r^2 = r_\perp^2 + r_\varparallel^2$. Each row is a side-by-side comparison between two HOD models with symmetric changes in the bias parameter, as defined in Table. \ref{tab:hod_param}. There are substantial changes on small scales in may cases, and the finger-of-god effect is especially exacerbated.
	
	For assembly bias in rows 2 and 3, because re-assigning halo masses in our implementation essentially changes the mass distribution of galaxies, reversing the sign of the assembly bias parameter does not simply result in the opposite change in the correlation. With the exception of assembly bias models, symmetrical positive and negative changes in the HOD parameters in all the other bias models induce symmetrical $\Delta \xi$ when comparing columns 1 to 3, or columns 2 to 4. All single-variation models tested produce distinct $\Delta\xi$ patterns (up to normalisation by $r^2$) on small scales and can be easily distinguishable from each other. When more than source of galaxy bias is present, the resulting 2D correlation will be a combination of all the contributions, making the pattern difficult to parse and likely creating degeneracies. Comparing the pre- and post-reconstruction columns, we see that the peripheral regions become noisy after reconstruction. This means that although the decorations imposed may cause nontrivial changes in 2PCF on intermediate scales around $40$, these changes are largely removed by reconstruction. 
	
	% delta xi0, xi2 for all models
	\begin{figure*} 
		\includegraphics[width=\linewidth]{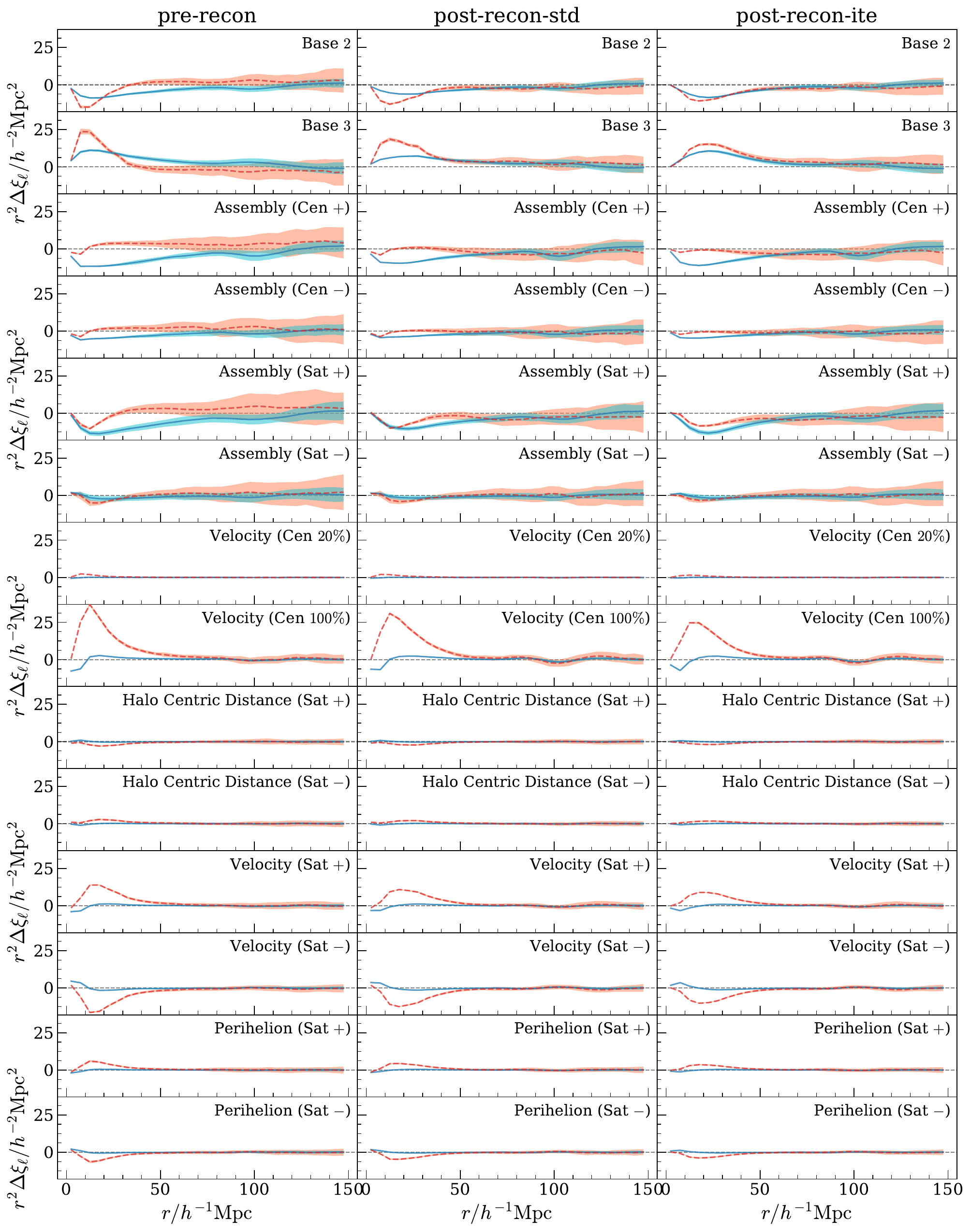}
		\caption{
			(Colour online) Phase-matched difference of jackknife multipoles for each biased HOD model with respect to the baseline; monopole in blue solid line, quadrupole in red dashed line. Every row is a biased HOD model with and without reconstruction applied. The shaded regions are ten times the size of the actual $1\sigma$ error bands, which would have been hardly visible. All axes share the same scaling.}
		\label{fig:delta_xi}
	\end{figure*}
	Fig. \ref{fig:delta_xi} shows the phase-matched differences of the first two jackknife multipoles for each biased HOD model with respect to the baseline model, i.e. $\Delta \xi_\ell = \xi_{\ell,\text{model}} - \xi_{\ell,\text{base}}$ rescaled by $r^2$. The sample variance across simulation boxes is already very small after jackknife resampling, and the $1\sigma$ errors are too small when plotted, so the error regions in Fig. \ref{fig:delta_xi} are artificially enlarged 10 times to improve visibility and demonstrate the reduction of sample variance by two reconstruction methods.
	
	The correlation multipoles show the same trend as does the 2D redshift-space correlation function. Satellite variations in the first two rows change small-scale clustering significantly, indicating the re-distribution of satellite galaxies mostly changing the 1-halo term; on larger scales the changes get noisier but one can barely see that there exist rises and drops around the BAO scale, and monopole and quadrupoles change in opposite directions. The assembly bias models in rows 3-6 re-assign halo masses by ranking halo pseudomasses, and the mass distribution of halos selected with the mass cut varies across simulation boxes, so these plots are the noisiest ones; the changes in 2PCF are not symmetrical as the assembly bias parameters flips signs. From the first six rows, one notices that reconstruction shrinks the error bands considerably, reducing the sample variance. It also reduces the net change, bringing the multipoles closer to the unbiased zero point.
	 
	As velocity dispersion of central galaxies is increased from 20\% to 100\%, the multipole differences grow drastically, and a decrease in BAO peak amplitude is seen on large scales after reconstruction right around 100\hmpc, as central velocity bias smears the central-central contribution in the 2-halo term. Models in the last six rows incorporate sub-halo-scale physics and only significantly affect small-scale clustering up to about 50\hmpc as expected. These models are extremely stable across random realisations, with very small scatter in spite of highly exaggerated error ranges.	One can see that for each pair of models with symmetric changes in the HOD parameters, the $\Delta\xi_\ell$ plots are essentially symmetric in the same way as in the redshift-space 2PCF (Fig. \ref{fig:delta_rppi}). Again, the difference between standard and iterative reconstruction mostly lies in small to intermediate scales, and on large scales they are visually the same.
	
	\subsection{Effects of Galaxy Bias on the Acoustic Scale Measurement}
	\label{sec:alpha_biases}
	
	%		\begin{figure}
	%			\includegraphics[width=\columnwidth]{fig_baofit_chisq.pdf}
	%			\caption{
	%				(Colour online) The $\chi^2$ grids in the anisotropic BAO scale $(\alpha_\perp, \alpha_\varparallel)$ space from fitting one jackknife sample with the fiducial fitting model (\textit{poly}3 form of nuisance polynomial, $50 h^{-1}\si{\mega\parsec}$ to $150 h^{-1}\si{\mega\parsec}$ fitting range). The best-fit $(\alpha_\perp, \alpha_\varparallel)$ positions with minimum $\chi^2$ values are marked by crosses; ellipses surrounding the crosses are constant $\Delta\chi^2$ contours representing $1\sigma, 2\sigma, 3\sigma$, etc. confidence regions. Iterative reconstruction makes the minimum sharper and lower than standard reconstruction, reducing the best-fit $\chi_\text{min}^2$ value almost by half. todo: expand range so 2sigma can be seen in full}
	%			\label{fig:chisq_grid}
	%		\end{figure}
	
	%pre-recon fitting is not robust against galaxy biases.
	%With the fiducial fitting model (\textit{poly}3 form of nuisance polynomial, $50 h^{-1}\si{\mega\parsec}$ to $150 h^{-1}\si{\mega\parsec}$ fitting range) applied to all pre- and post-reconstruction samples, iterative reconstruction consistently yields sharper $\chi^2$ minima and lower best-fit $\chi^2_\text{min}$ values; there is a negligible difference in the BAO $\alpha$ values between standard and iterative reconstructions. An example of this is shown in Fig. \ref{fig:chisq_grid} for a jackknife sample.
	
	\begin{table*}
		\caption{BAO fitting results for all biased HOD models, subtracted by the corresponding $\alpha$ of the baseline case. Values are the mean and jackknife-corrected $1\sigma$ uncertainty of all simulation boxes. The two post-recon columns in the first header row represent standard and iterative reconstructions respectively.}
		\label{tab:bao_results}
		\begin{tabular}{
			l
			S[table-format=-2.2(2)]
			S[table-format=-2.2(2)]
			S[table-format=-2.2(2)]
			S[table-format=-2.2(2)]
			S[table-format=-2.2(2)]
			S[table-format=-2.2(2)]
			}
			\hline
			\hline
			Reconstruction Type
			&\multicolumn{2}{c}{pre-recon}
			&\multicolumn{2}{c}{post-recon-std}
			&\multicolumn{2}{c}{post-recon-ite}\\
			BAO Direction
			& $\Delta\alpha_\perp/\%$ & $\Delta\alpha_\varparallel/\%$
			& $\Delta\alpha_\perp/\%$ & $\Delta\alpha_\varparallel/\%$
			& $\Delta\alpha_\perp/\%$ & $\Delta\alpha_\varparallel/\%$\\
			\hline
			Base 2
			&-0.36 (07) & 0.22 (12)
			&-0.13 (05) & 0.38 (08)
			&-0.09 (05) & 0.31 (07)\\
			Base 3
			& 0.36 (10) &-0.35 (14)
			& 0.05 (06) &-0.15 (06)
			& 0.01 (05) &-0.02 (07)\\
			Assembly Bias (Centrals $+$)
			& 0.12 (11) &-0.43 (21)
			& 0.06 (07) & 0.11 (11)
			& 0.04 (07) & 0.13 (11)\\
			Assembly Bias (Centrals $-$)
			&-0.16 (12) & 0.25 (24)
			&-0.21 (08) & 0.14 (13)
			&-0.15 (08) & 0.10 (14)\\
			Assembly Bias (Satellites $+$)
			&-0.18 (15) & 0.20 (19)
			& 0.07 (09) & 0.19 (11)
			& 0.12 (07) & 0.12 (09)\\
			Assembly Bias (Satellites $-$)
			&-0.38 (16) & 0.05 (22)
			&-0.21 (08) & 0.21 (07)
			&-0.20 (07) & 0.15 (09)\\
			Velocity Bias (Centrals 20\%)
			& 0.01 (03) &-0.00 (05)
			&-0.00 (03) &-0.08 (03)
			& 0.00 (02) &-0.06 (04)\\
			Velocity Bias (Centrals 100\%)
			& 0.11 (06) &-0.59 (14)
			& 0.16 (06) &-0.66 (28)
			& 0.18 (03) &-0.73 (13)\\
			Halo Centric Distance Bias (Satellites $+$)
			&-0.05 (03) &-0.01 (05)
			&-0.04 (03) &-0.03 (04)
			&-0.03 (03) & 0.00 (06)\\
			Halo Centric Distance Bias (Satellites $-$)
			&-0.03 (04) & 0.04 (07)
			& 0.01 (02) &-0.02 (04)
			& 0.01 (03) &-0.02 (04)\\
			Velocity Bias (Satellites $+$)
			& 0.02 (04) &-0.14 (10)
			& 0.02 (04) &-0.16 (06)
			& 0.03 (03) &-0.15 (05)\\
			Velocity Bias (Satellites $-$)
			&-0.05 (04) & 0.16 (07)
			&-0.02 (03) & 0.12 (05)
			&-0.02 (03) & 0.16 (06)\\
			Perihelion Distance Bias (Satellites $+$)
			& 0.00 (04) &-0.09 (07)
			&-0.01 (03) &-0.11 (05)
			&-0.00 (03) &-0.09 (06)\\
			Perihelion Distance Bias (Satellites $-$)
			&-0.02 (04) & 0.12 (06)
			&-0.02 (02) & 0.07 (05)
			&-0.02 (02) & 0.10 (04)\\
			\hline
		\end{tabular}
	\end{table*}
	
	\begin{figure*}
		\includegraphics[width=0.68\linewidth]{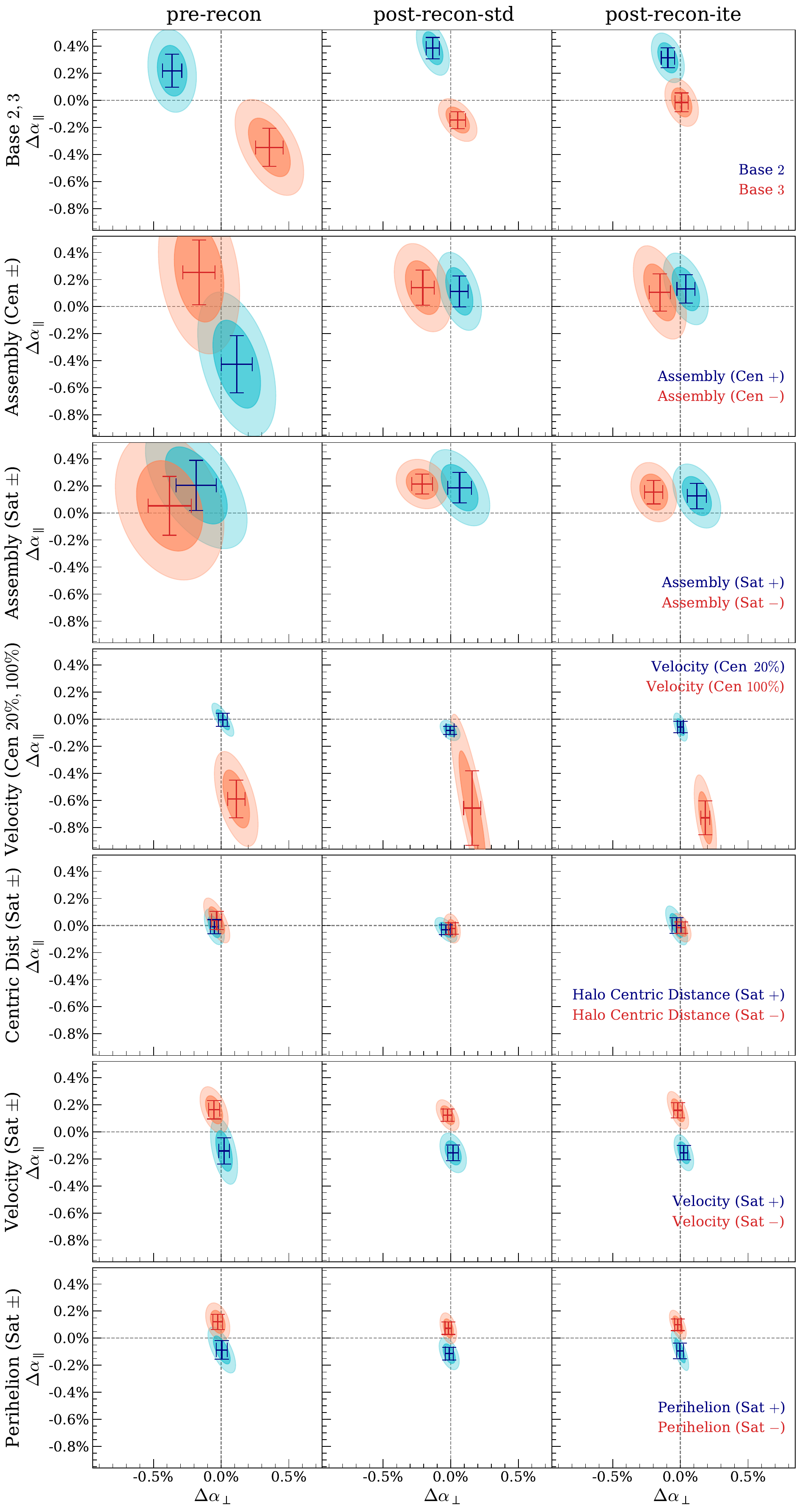}
		\caption{
			(Colour online) Phase-matched difference between the BAO scales found in biased HOD models and the baseline model, $\Delta\alpha = \alpha_\text{model} - \alpha_\perp\text{baseline}$ in both directions. Every pair of comparable models are plotted in the same row. Each mean and uncertainty patch is derived from $N_\text{bos} = 36$ fitting results.}
		\label{fig:delta_alpha}
	\end{figure*}
	
	Having computed	the correlation functions of our various models, we fit the acoustic signal to yield transverse and radial scale measurements. Then the $\alpha$ values are subtracted by the baseline model value for each simulation box in a phased-matched manner. The subtraction yields the differential change in BAO $\alpha$ for each biased model relative to the baseline. Phase-matching is critical when taking the difference, because every box has its own sample variance shared by the realisations of all models and this sample variance can be cancelled out effectively as demonstrated in \textsection\ref{sec:bao_fitting}. Table \ref{tab:bao_results} summarises the shifts of the acoustic scale for all biased HOD models with respect to the baseline model. Two reconstruction methods are listed separately. The same numerical results are plotted in Fig. \ref{fig:delta_alpha}, where each row shows a pair of comparable models and each column is a pre- or post-reconstruction sample type.
	
	The first two models in the first row with opposite variations in satellite mass cutoff and power law exponent see asymmetric shifts. It is encouraging to see that iterative reconstruction manages to completely eliminate the bias of the Base 3 model (higher $M_1$ and $\alpha$), bringing the red cross back right onto the origin, and slightly reduces the bias of the Base 2 model (lower $M_1$ and $\alpha$). This is one of the only two cases where iterative reconstruction noticeably performs better than standard reconstruction at restoring the unbiased BAO scale. By lowering  $M_1$ and $\alpha$, Base 2 effectively moves satellite galaxies from high mass halos to lower mass halos. There are many more possible locations to put them at a lower mass cut, and low mass halos do not mark the highest initial overdensities as much as high mass halos do, so correcting for the bias for Base 2 is naturally more difficult than for Base 3, resulting in a residual shift of 0.3\% post-reconstruction primarily in the radial direction.
	
	For the assembly bias models in rows 2 and 3 of Fig. \ref{fig:delta_alpha}, the shifts induced by opposite bias parameters are almost symmetrical in the horizontal, or traverse, direction, but in the radial direction all models are slightly biased toward a higher $\alpha_\varparallel$ regardless of the sign of the bias parameter. The previous plots (Fig. \ref{fig:delta_rppi} and \ref{fig:delta_xi}) do not clearly separate the two directions and show this difference. The second and third rows have similar mean values for the same coloured models post-reconstruction, only the uncertainty is smaller for satellite assembly bias. This resemblance  implies that, whether assembly bias is present in the central or satellite galaxies, the end effect on the acoustic scale is virtually the same if reconstruction is applied. In other words, assembly bias does not distinguish between centrals and satellites. Comparing the models with negative signs only (red markers and orange ellipses), however, one sees that the satellite assembly bias results are better constrained than central assembly bias ones, making the same 0.2\% shift more significant for the satellite negative model. But it is only about $2\sigma$, and at the 0.2\% level all four models are consistent with zero shift.
	
	For the velocity bias models in the fourth row, a realistic dispersion at the 20\% level hardly causes any shift, whereas the more extreme 100\% case resulted in large shifts of 0.7\%. This is the other one of the two cases where iterative reconstruction performs noticeably better than standard reconstruction. In fact, there is even a qualitative difference here\textemdash the orange error region of standard reconstruction seems to be consistent with zero shift at the $2\sigma$ level, but the iterative reconstruction uncertainty is much more confined and certainly rules out the no shift possibility.
	
	The last three rows are six models with sub-halo-scale astrophysics. Galaxy bias originating from halo centric distances of DM particles clearly does not impact the BAO measurement. On the other hand, satellite velocity bias and perihelion bias (or dependence on the total mechanical energy of DM particle) do make small differences of 0.1\% to 0.2\%. Satellite velocity bias results in almost $0.2\%$ shifts with just above $2\sigma$ significance. The perihelion bias models have smaller shifts of 0.1\%, which are insignificant. The symmetric pattern of shifts can again be seen in the last two rows for models of opposite signs. Again, at the 0.2\% level all six models are consistent with zero shift.
	
	Iterative reconstruction offers better BAO fitting results than standard reconstruction in general. In rows 1 and 4 of Fig. \ref{fig:delta_alpha}, it made a real difference by significantly reducing the bias and uncertainty in the acoustic scale measurement, as mentioned above. For other models, it is not drastically better and the mean $\alpha$ values are extremely close to the standard reconstruction ones, but still it yields slightly less bias in the acoustic scale and tighter constraints.
	
	Lastly, we also notice a reliable inverse correlation between the sign of $\Delta\alpha_\varparallel$ and the sign of $\Delta \xi_2$, when comparing Fig. \ref{fig:delta_alpha} with \ref{fig:delta_xi}. This is obeyed by every model tested, even if the changes in 2PCF and acoustic scale are not very significant. A positive $\Delta \xi_2$ relative to the baseline corresponds to a negative $\Delta\alpha_\varparallel$, and vice versa. While we used the fiducial BAO fitting model with a fitting range of $r \in (50\hmpc, 150\hmpc)$ and the $\textit{poly}3^\prime$ nuisance form, improved fitting methods may exploit this inverse correlation by extending the range to smaller scales.
	
	\section{Conclusions}
	\label{sec:conclusions}
	
	In this work, we test the effect of galaxy bias on the acoustic scale by considering several bias mechanisms to their extremes. With accurate N-body simulations in a total volume of $48 \, h^{-3}\si{\giga\parsec\tothe{3}}$ and a generalised HOD approach, every biased model can be compared to the baseline to derive the differential shift in the acoustic scale measurement that precisely corresponds to the change in the input HOD parameters.
	
	We find a 0.3\% shift in the line-of-sight acoustic scale for one variation in the satellite galaxy population, the Base 2 model. The model with an extreme level of velocity bias of the central galaxies produces the largest shift in the BAO measurement, $0.7\%$ relative to the unbiased scale. All the other bias models result in either small ($0.2\%$ or less) or statistically insignificant ($2\sigma$ or less) shifts. Except for the highly unlikely event that the central galaxies have very large velocity dispersions relative to the halo bulk (close to the typical speed of DM particles), we find the shifts caused by single-variation models to be below 0.3\%. However, this is by no means a claim that the theoretical systematic error in the acoustic scale measurement due to galaxy bias originating from the halo-galaxy connection in the halo model is \textit{only} $0.3\%$. Observed galaxy samples from redshift surveys may well be subject to not one, but many processes which can introduce galaxy bias. Combinations of bias mechanisms at play may potentially compound the uncertainties we found and push the BAO shift above 0.3\%.
	
	That said, the biggest shift of the acoustic scale at 0.7\% comes from increased velocity bias for central galaxies, which seems readily detectable\textemdash there would be sizeable changes in the velocity dispersion of clusters compared to their weak lensing masses. In addition, these bias models also create substantial changes on small scales, which may in fact allow one to detect these effects and thereby improve the modelling of the acoustic scale. In Fig. \ref{fig:delta_rppi}, many bias models have greatly increased finger-of-god effect in the correlation function, both pre- and post-reconstruction. Given the inverse correlation we found between the sign of $\Delta\alpha_\varparallel$ and the sign of $\Delta \xi_2$, small-scale clustering data offer a promising opportunity to correct for galaxy bias and further calibrate the acoustic peak ruler with future development in the 2PCF fitting formalism.

	In regards to reconstruction, both standard and iterative reconstruction methods show similar efficacy in reducing the imposed bias and recovering the unbiased 2PCF and acoustic scale. In terms of the performance for BAO fitting, iterative reconstruction is more robust against galaxy bias, bringing the BAO measurements closer to the true, unbiased acoustic scale; it is also more precise, being less prone to sample variance and producing less uncertainty. It is a promising new method with the potential to benefit from further development and optimisation in the future. Although it currently requires an order of magnitude more resources in CPU time and memory allocation than standard reconstruction does, reconstruction and BAO fitting comprise a minor fraction of the computation time in comparison to the time needed for genuine $N$-body simulations. The extra time needed for iterative reconstruction is worthwhile If one is concerned with minimising the bias and uncertainty in the BAO analysis.
	
	While current galaxy surveys, such as the Sloan Digital Sky Survey
	(SDSS) III Baryon Oscillations Spectroscopic Survey (BOSS), measure the acoustic scale to about 1\% precision and are insensitive to the galaxy bias effects shown, upcoming dark energy experiments, including DESI, Euclid, and WFIRST, will make use of the BAO standard ruler to sub-percent level precision and these effects can no longer be overlooked. Our ability to account for or even correct for galaxy bias in the modelling of the acoustic scale directly impacts the measurement precision of the BAO ruler and the success of upcoming surveys. Having shown that the systematic effects of galaxy bias alone could amount to 0.3\%, we find the majority of the shifts values in Table \ref{tab:bao_results} encouraging and note that these bias effects may be detected. Accurate modelling of the galaxy-halo connection, in conjunction with the bias that comes with it, is a need of growing urgency as the statistical uncertainties of larger surveys approach the level of cosmic variance limit and theoretical systematics. Future analysis of the BAO signal may benefit from the inclusion of velocity dispersion and small-scale clustering to mitigate the non-negligible systematic effects of galaxy bias.
	
	\section*{Acknowledgements}
	
	The authors would like to thank Lehman Garrison for producing the \abacus simulation products and managing the computing cluster. DYT thanks Lehman Garrison for helpful discussions on Corrfunc and \rockstar halo and particle catalogue products, Ryuichiro Hada for providing the reconstruction code, Ashley Ross for sharing his BAO fitter used in BOSS DR12, and Prof. Steve Ahlen for his continued support and mentorship. DJE is supported by U.S. Department of Energy grant DE-SC0013718 and as a Simons Foundation Investigator. DYT is supported by U.S. Department of Energy grant DE-SC0015628.
	
	
	\bibliographystyle{mnras}
	\interlinepenalty=10000
	\bibliography{paper} % if your bibtex file is called example.bib
	
	%%%%%%%%%%%%%%%%% APPENDICES %%%%%%%%%%%%%%%%%%%%%
	
	\appendix

	\bsp	% typesetting comment
	\label{lastpage}
\end{document}